\documentclass[%
 reprint,
 superscriptaddress,
 amsmath,amssymb,
 aps,
]{revtex4-2}

\usepackage{graphicx}% Include figure files
\usepackage{dcolumn}% Align table columns on decimal point
\usepackage{bm}% bold math
\usepackage{mhchem}
\usepackage{multirow}
\usepackage{xcolor}
\usepackage[utf8]{inputenc}
\usepackage{amsmath,amssymb}
\newcommand{\recheck}[1]{{\color{black}#1}}

\begin{document}

\preprint{APS/123-QED}

\title{Pretraining of attention-based deep learning potential model for molecular simulation}

\author{Duo Zhang}
\affiliation{AI for Science Institute, Beijing 100080, P. R.~China\\}
\affiliation{DP Technology, Beijing 100080, P. R.~ China\\}
\affiliation{Academy for Advanced Interdisciplinary Studies, Peking University, Beijing 100871, P. R.~China\\}

\author{Hangrui Bi}
\affiliation{AI for Science Institute, Beijing 100080, P. R.~China\\}
\affiliation{DP Technology, Beijing 100080, P. R.~ China\\}

\author{Fu-Zhi Dai} 
\affiliation{AI for Science Institute, Beijing 100080, P. R.~China\\}

\author{Wanrun Jiang}
\affiliation{AI for Science Institute, Beijing 100080, P. R.~China\\}

\author{Linfeng Zhang}
 \email{linfeng.zhang.zlf@gmail.com}
\affiliation{AI for Science Institute, Beijing 100080, P. R.~China\\}
\affiliation{DP Technology, Beijing 100080, P. R.~ China\\}

\author{Han Wang}
 \email{wang\_han@iapcm.ac.cn}
\affiliation{Laboratory of Computational Physics, Institute of Applied Physics and Computational Mathematics, Beijing 100094, P. R.~China\\}
\affiliation{HEDPS, CAPT, College of Engineering, Peking University, Beijing 100871, P. R.~China\\}

\begin{abstract}
Machine learning assisted modeling of the inter-atomic potential energy surface (PES) is revolutionizing the field of molecular simulation.
With the accumulation of high-quality electronic structure data, a model that can be pretrained on all available data and finetuned on downstream tasks with a small additional effort would bring the field to a new stage.
Here we propose DPA-1, a Deep Potential model with a novel attention mechanism, which is highly effective for representing the conformation and chemical spaces of atomic systems and learning the PES. 
We tested DPA-1 on a number of systems and observed superior performance compared with existing benchmarks.
When pretrained on large-scale datasets containing 56 elements, DPA-1 can be successfully applied to various downstream tasks with a great improvement of sample efficiency.
Surprisingly, for different elements, the learned type embedding parameters form a $spiral$ in the latent space and have a natural correspondence with their positions on the periodic table, showing interesting interpretability of the pretrained DPA-1 model. 
\end{abstract}

\maketitle

%\tableofcontents

\section{Introduction}
Reliably representing the inter-atomic potential energy surface (PES) is core to the study of properties of molecules and materials in computational physics, chemistry, materials science, biology, etc.
%The energy function of a system not only determines its state distribution at thermo-equilibrium but also gives the force field needed for simulating its dynamics. 
While electronic structure methods typically give accurate and transferable PES, they are prohibitively expensive for scaling to systems of more than thousands of atoms.
On the other hand, empirical force fields 
%based on simple analytical functions of atomic coordinates 
are much more efficient, but are inherently limited by their accuracy in many applications.
%With the rapid development of machine learning (ML), 
By properly integrating machine learning (ML) methodologies and physical requirements like extensiveness and symmetries, 
various methods have emerged to address the accuracy $v.s.$ efficiency dilemma in the realm of PES modeling~\cite{ behler2007generalized, bartok2010gaussian,thompson2015spectral, gilmer2017neural, schutt2017schnet, zhang2018deep,zhang2018end, drautz2019atomic,gasteiger2019directional, zhang2019embedded, gasteiger2021gemnet}.
%. The proposed ML-based methods vary from shallow neural networks and kernel-based approaches \cite{blank1995neural, handley2009optimal, behler2007generalized, bartok2010gaussian} to more recent methods based on deep learning, in particular, local descriptor based methods \cite{bartok2010gaussian, thompson2015spectral, zhang2018deep, drautz2019atomic} and message passing neural networks (MPNN)\cite{gilmer2017neural, schutt2017schnet, gasteiger2019directional, gasteiger2021gemnet}.
Arguably, %the field of molecular dynamics simulation is heading towards 
a new paradigm is forming: electronic structure methods are no longer used to generate the driving forces during molecular dynamics simulations but are used to generate data for training their alternatives, ML-based PES models.

Despite remarkable achievements of ML-based PES models~\cite{deringer2021gaussian,unke2021machine,wen2022deep}, challenges still remain.
For a domain expert who would like to apply such methodologies in their applications, a natural first question is on the efforts 
%like the amount of training data 
needed for obtaining a reliable PES model:
Are there ready-to-use PES models?
If not, what would be the amount of training data and time cost required?
Can we take advantage of the ever-increasing publicly-available training data?

To address these issues, there have been several efforts. % falling into two categories. 
On one hand, general-purpose models for various systems, such as silicon~\cite{Bartok2018Silicon},
phosphorus~\cite{deringer2020general},
water~\cite{Zhang_2021_prl}, metals and alloys~\cite{jiang2021accurate,Szlachta_2014_prb,wang2022tungsten,Wang_2021_msmse,Wen_2021_npj}, etc., have been developed and are directly applicable to relevant studies.
However, the range of applicability of such models is typically limited to small conformation or chemical space.
For example, for alloys, the majority of general-purpose ML models are developed for systems with at most two element types.
On the other hand, several efficient data generation protocols have been developed~\cite{podryabinkin2017active, smith2018less, zhang2019active, zhang2020dp}, of which a representative is DP-GEN~\cite{zhang2019active,zhang2020dp}, a concurrent learning procedure that iteratively explores the configuration space using models trained with existing data, and then labels only those configurations with high uncertainty level.
Even with these protocols, the computational effort needed for complicated systems is still prohibitive.
For example, to train a fairly general-purpose model for the AlMgCu alloy system, 100k density functional theory (DFT)~\cite{kohn1965self, car1985unified} calculations were ultimately performed, resulting in the cost of ten million CPU core hours~\cite{jiang2021accurate}.

With the accumulation of high-quality electronic structure data covering almost all the elements on the periodic table, it is becoming possible to systematically develop pretraining schemes, which have been widely adopted in areas like computer vision (CV)~\cite{russakovsky2015imagenet, dosovitskiy2020image} and natural language processing (NLP)~\cite{devlin2018bert, brown2020language}. %and take the field of molecular simulation to a new stage.
In these schemes, one first trains a unified model on large-scale datasets and then finetunes it for downstream tasks, 
%using techniques like transfer learning
expecting that a good representation can be learned in the first stage, and the amount of supervised data needed for the second stage will be significantly reduced.
  \recheck{Recently, the pretraining-finetuning idea is applied to organic molecules systems for the energy and force predictions~\cite{smith2017ani,smith2019approaching}, and to tackle tasks beyond representing the PES~\cite{liu2022pretraining,stark2021_3dinfomax,zhou2022uni}}.
%Such a new possibility also brings new requirements on the PES model architecture. 
Unfortunately, most ML-based PES models are premature for such schemes at scale \recheck{in materials applications}. 
Taking the widely used two versions of Deep Potential models~\cite{zhang2018deep,zhang2018end} as examples, 
the ML parameters are element-type-dependent, making it highly inefficient when the training data containing many elements.

  \recheck{
Constant efforts have been devoted to adapt the architecture of the ML-based PES models for large datasets. 
Among them, one class of models named equivariant graph neural networks (GNN)~\cite{thomas2018tensor} that is built upon convolutions over atomic graphs of node and edge equivariant representations has shown promise of training on large datasets.
SchNet~\cite{schutt2017schnet}, PaiNN~\cite{schütt2021equivariant}, GemNet-OC~\cite{gasteiger2022graph}, DimeNet++~\cite{gasteiger2022fast}, PFP~\cite{takamoto2022towards}, SCN~\cite{zitnick2022spherical}, SpinConv~\cite{shuaibi2021rotation} and Equiformer/EquiformerV2~\cite{liao2023equiformer, liao2023equiformerv2} are trained on the OC20/OC2M~\cite{chanussot2021open} dataset containing about 133M/2M data frames of 56 elements. 
These models are benchmarked by the accuracy of energy, force and stable structure predictions. 
Very recently, it has been shown that introducing the attention architecture~\cite{liao2023equiformer} in a GNN model improves the performance on the OC20/OC2M dataset~\cite{liao2023equiformerv2}.
Chen and Ong~\cite{chen2022universal} proposed M3GNet, which was able to train on a subset of the Materials Project~\cite{jain2013commentary} that contains 187,687 configurations encompassing 89 elements and labeled at the generalized gradient approximation (GGA)~\cite{perdew1996generalized} or GGA+U level.
Takamota et.~al.~\cite{takamoto2022towards} introduced the PFP model, which was trained on a dataset composed of molecular and crystal configurations including approximately $ 9\times 10^6$ frames of 45 elements. 
Choudhary et.~al.~\cite{choudhary2023unified} developed the ALIGNN model, and they were able to train the model on a subset of the JARVIS-DFT dataset~\cite{choudhary2020joint} that is composed of 307,113 data frames of 89 elements. 
The M3GNet, PFP and ALIGNN models are proposed as ``universal'' potential models, however, their accuracies are not on-par with PES models trained for a specific materials applications. 
}

%   \recheck{
% Models such as PaiNN~\cite{schütt2021equivariant}, EQGAT~\cite{le2022equivariant}， GemNet-OC~\cite{gasteiger2022graph}, Equiformer~\cite{liao2023equiformer}, M3GNet~\cite{chen2022universal}, and others have been developed to train on large datasets like the Open Catalyst 2020 Dataset (OC20)~\cite{chanussot2021open} or the Materials Project~\cite{jain2013commentary}, which contain dozens of elements. These models have achieved impressive accuracy by incorporating novel structures such as graph-based attention mechanisms~\cite{vaswani2017attention, dosovitskiy2020image}, thus demonstrating greater generalization ability. They all aim to develop ``universal" PES models. However, for real-world material applications, most of these methods face the following challenges:}

  \recheck{
The equivariant GNN models are potential candidates for pretraining, several issues worth special attention before applying them in down-stream real-world applications. 
First, the GNN approaches are not well-suited for massively parallel molecular dynamics simulations~\cite{musaelian2022learning}. 
The update of each GNN layer requires communications between spatially decomposed sub-regions of the system. 
In each evaluation of the energy and forces, in total several to a dozen of such updates are required, which may lead to a substantial communication overhead in massively parallel high performance super-computers.
Second, some models, such as PaiNN, GemNet-OC, SCN, Equiformer/EquiformerV2, directly predict forces using rotationally equivariant networks~\cite{schütt2021equivariant, le2022equivariant, gasteiger2022graph, liao2023equiformer} instead of energy gradients with respect to atomic coordinates. 
Therefore, the predicted force is not conservative, which serves as a basic assumption in guaranteeing the accuracy of molecular simulations~\cite{bond2007molecular}. 
% This can lead to nonconservation of energy, potential collapse in molecular dynamics simulations, and overfitting in training on real-world materials systems.
Last but not least, some models, such as GemNet-OC, SpinConv, M3GNet, ALIGNN are not smooth, i.e. a sudden energy jump may happen as the positions of atoms infinitesimally varies. 
This leads to a non-conserved energy in the Hamiltonian dynamics simulations, which is used in computing the dynamical properties like diffusion constant and viscosity. 
}

%   \recheck{Therefore, the performance of these ``universal" PES models on real-life tasks and large-scale applications, both from scratch or utilizing pretraining and then fine-tuning on downstream tasks, remains to be seen.}

%   \recheck{Recently, efforts have been made to apply the pretraining scheme to organic molecules~\cite{smith2017ani,smith2019approaching} and consider tasks beyond representing the PES~\cite{liu2022pretraining,stark2021_3dinfomax,zhou2022uni}. In this work, we focus more on extended systems and aim for a much larger elemental space. Very recently, methods such as MEKRR~\cite{falk2023transfer} have also begun to explore similar topics. They used pre-trained GNN representations from the OC20 dataset and tested their transferability of energies on some downstream tasks. However, their scalability to larger downstream datasets still needs improvement.}

\recheck{
By far, how much may the downstream materials applications benefit from the ML models trained on the large-scale datasets are still not clear.}
\recheck{To answer the question,} in this article, we propose DPA-1, 
a Deep Potential model with a novel attention mechanism.
% , which   \recheck{use local desccriptor, which is highly suitable for paralell simulations on large-scale systems containing millions of atoms~\cite{jia2020pushing}, conserves force, maintains smoothness and} is highly effective for learning the inter-atomic interactions and, upon pretraining, can significantly reduce the additional efforts for downstream tasks.
  \recheck{Designed with a local descriptor, this model is exceptionally well-suited for parallel simulations on large-scale systems containing millions of atoms~\cite{jia2020pushing}. Notably, DPA-1 predicts conservative forces, ensures smoothness, and demonstrates outstanding efficacy in learning inter-atomic interactions. Moreover, once pretrained, DPA-1 can significantly decrease the supplementary efforts needed for subsequent downstream tasks.}
We tested DPA-1 on various systems and observed superior performance compared with existing benchmarks.
Then we took AlMgCu alloy systems~\cite{jiang2021accurate} as an example, showing that after pretraining with single-element and binary samples, DPA-1 can save around 90\% ternary samples compared with the DeepPot-SE model~\cite{zhang2018end}.
Finally, we pretrained DPA-1 using the OC20 dataset, which consists of 56 elements, and successfully applied it to various downstream tasks.
We checked the interpretability of the pretrained model by looking into the learned embedding parameters for different element types, 
finding that the 56 elements are arranged on a $spiral$ in the latent space, which has a natural correspondence with their physical properties on the periodic table.
Above all, we believe that DPA-1 and the pretraining scheme will bring the field of molecular simulation to a new stage.

\section{Method}
Consider a system of $N$ atoms, the elemental types are $\mathcal{A} = \left\{\alpha_1,\alpha_2,...,\alpha_i,...,\alpha_N \right\}$, and the atomic coordinates are $\mathcal{R}=\left\{\boldsymbol{r}_1,\boldsymbol{r}_2,...,\boldsymbol{r}_i,...,\boldsymbol{r}_N\right\}$, with $\boldsymbol{r}_i$ being the three Cartesian coordinates of atom $i$.
The PES of the system is denoted by $E$, a function of elemental types and coordinates, i.e. $E=E(\mathcal{A}, \mathcal{R})$.
For each atom $i$, consider its neighbors $\{j|j\in{\mathcal{N}_{r_c}(i)}\}$,
where $\mathcal{N}_{r_c}(i)$ denotes the set of atom indices $j$ such that $r_{ji} < r_c$,
with $r_{ji}$ being the Euclidean distance between atoms $i$ and $j$.
$E$ is represented as the summation of atomic energies $\left\{e_1, e_2, ..., e_i, ..., e_N\right\}$,
%Without loss of generality, we can omit the long-range interactions and assume 
where the atomic energy $e_i$ only depends on the
information of ${\mathcal{N}_{r_c}(i)}$.
%neighboring environment within a sphere of radius $r_c$ as the cut-off. 
We define $N_i=|\mathcal{N}_{r_c}(i)|$, the cardinality of the set $\mathcal{N}_{r_c}(i)$.
We use $\mathcal{A}^i$ to denote element types in ${\mathcal{N}_{r_c}(i)}$, and $\mathcal{R}^{i}\in \mathbb{R}^{N_{i} \times 3}$ their corresponding coordinates relative to $i$. % of the neighbors. 
The atomic energy $e_i$ is thus a function of $\mathcal{A}^i$ and $\mathcal{R}^{i}$.
The atomic force on atom $i$, $\mathcal{F}_i$, is defined as the negative gradient of the total energy with respect to $i$'s coordinate:
\begin{equation}
\label{eq:force}
\mathcal{F}_{i}=-\nabla_{\boldsymbol{r}_{\boldsymbol{i}}} E.
\end{equation}

We refer to Ref.~\cite{zhang2018end} for a detailed discussion on several requirements on PES modeling.
In particular, the PES has to be invariant under translation, rotation, and permutation of the indices of atoms with the same element types. 

\begin{figure*}
    \includegraphics[width=0.8\textwidth]{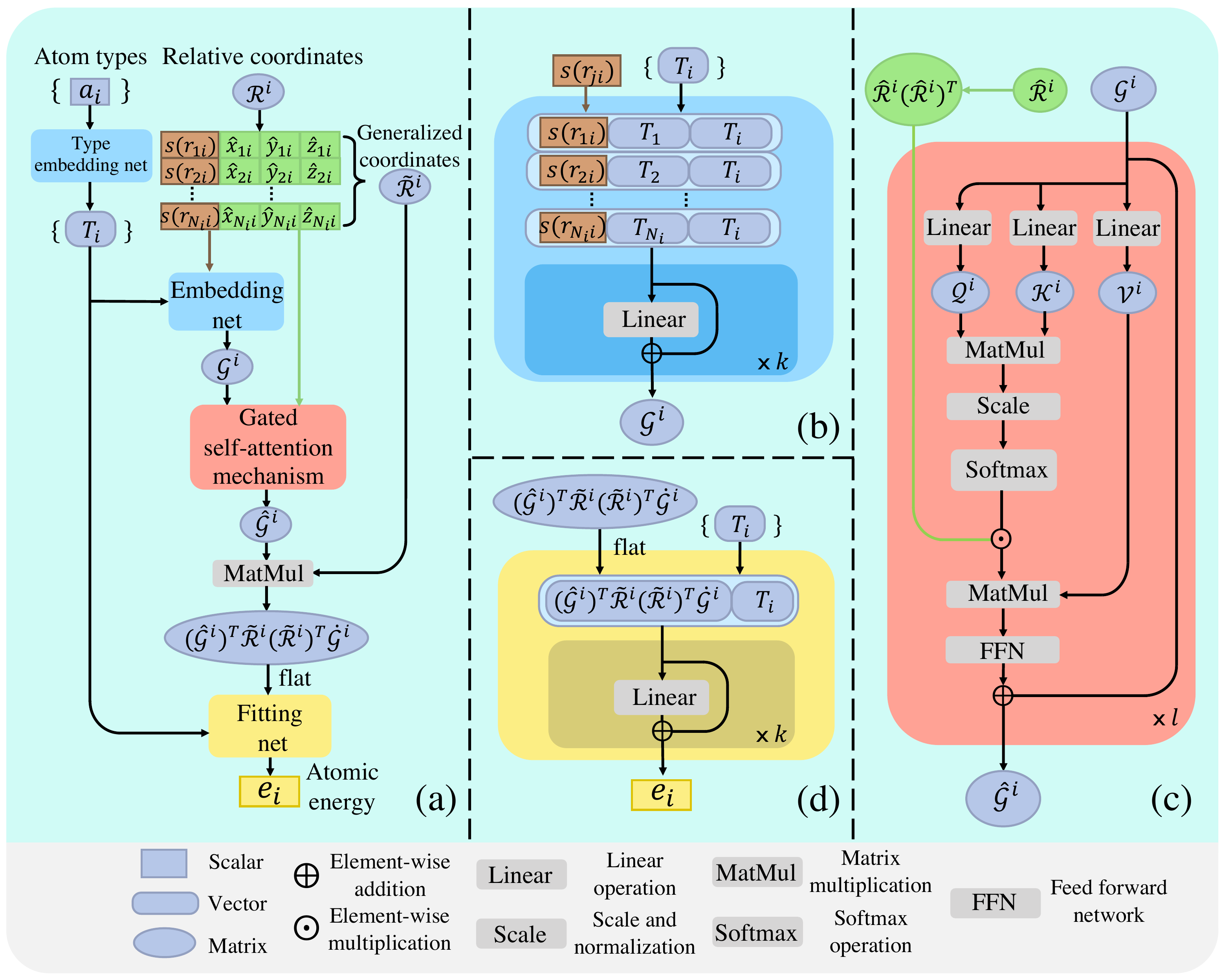}
    \centering
    \caption{Schematic illustration of DPA-1. 
    (a) Flowchart from $\mathcal{A}^i$  and $\mathcal{R}^{i}$ to the atomic energy $e_i$. 
    (b) Structure of the Embedding net, 
    which maps $s(r_{ji})$ and $T_{i}$, through multiple residual layers, to $\mathcal{G}^{i}$.
    (c) Self-attention mechanism on $\mathcal{G}^{i}$ through a standard scale-dot procedure gated by the angular information $\hat{\mathcal{R}}^{i}(\hat{\mathcal{R}}^{i})^{T}$.
    (d) Fitting net structures, similar to Embedding net, from the descriptors $\mathcal{D}^{i}$ and $T_{i}$ to final atomic energy $e_i$.
    }
    \label{fig:model}
\end{figure*}

The details of the model architecture are introduced below. 
We refer to Fig. \ref{fig:model} for the overall pipeline to predict the atomic energy $e_i$: from the embedded neighboring environment, through the self-attention scheme, to the symmetry-preserving descriptors, and finally to the fitting network. 

\label{sec:method}

\subsection{Local embedding matrix with type information}

We obtain the local embedding matrix with the following three steps. 
First, $\mathcal{R}^{i}$ is mapped to the generalized coordinates $\tilde{\mathcal{R}}^{i} \in \mathbb{R}^{N_{i} \times 4}$. 
In this mapping, each row of $\mathcal{R}^{i}, \left\{x_{ji}, y_{ji}, z_{ji}\right\}$, is transformed into a row of $\tilde{\mathcal{R}}^{i}$:
\begin{equation}
    \left\{x_{ji}, y_{ji}, z_{ji}\right\} \mapsto\left\{s\left(r_{ji}\right), \hat{x}_{ji}, \hat{y}_{ji}, \hat{z}_{ji}\right\},
\end{equation}
where $\left\{x_{ji}, y_{ji}, z_{ji}\right\} $ denotes the Cartesian coordinates of $\boldsymbol{r_{ji}} = \boldsymbol{r_j} - \boldsymbol{r_i}$, 
%and $r_{ji} = \vert \boldsymbol{r_{ji}} \vert$.
$\hat{x}_{ji}=\frac{s\left(r_{ji}\right) x_{ji}}{r_{ji}}, \hat{y}_{ji}=\frac{s\left(r_{j i}\right) y_{j i}}{r_{ji}}, \hat{z}_{ji}=\frac{s\left(r_{ji}\right) z_{ji}}{r_{ji}}$, and  $s\left(r_{j i}\right): \mathbb{R} \mapsto \mathbb{R}$ is a continuous and
differentiable scalar weighting function applied to each component, defined as:
\begin{widetext}
\begin{equation}
\label{eq:swf}
s\left(r_{j i}\right)= 
\begin{cases}
\frac{1}{r_{j i}} & r_{j i}<r_{c s} \\ 
\frac{1}{r_{j i}}\left[u^{3}\left(-6 u^{2}+15 u-10\right)+1\right] & r_{c s} \leq r_{j i}<r_{c}, \quad u=\frac{r_{j i}-r_{c s}}{r_{c}-r_{c s}} . \\ 
0 & r_{c} \leq r_{j i}
\end{cases}
\end{equation}
\end{widetext}
Here $r_{cs}$ is a smooth cutoff parameter that allows the components in $\tilde{\mathcal{R}}^{i}$ to smoothly go to zero at the boundary of the local region defined by $r_c$.
%, which is the smooth edition in original DeePMD.
% \LZ{the smoothing function has changed}

Second, we add the atomic type embedding as supplemental information. For atom $i$, the type embedding map $T_{i}$ is defined as:
\begin{equation}
\label{eq:tebd}
T_{i} = \phi_{T}(\alpha_{i}),
\end{equation}
where $\alpha_{i}$ is the atomic type of atom $i$ and $\phi_{T}$ is a one-hot-like embedding network mapping from $\alpha_{i}$ to a length-fixed vector.% as output.

Then, given both $\tilde{\mathcal{R}}^{i}$ and type embeddings $\left\{T_{i}\right\}\cup\left\{T_{j} \vert j\in{\mathcal{N}_{r_c}(i)}\right\}$, we define the local embedding matrix $\mathcal{G}^{i} \in \mathbb{R}^{N_{i} \times M_{1}}$:
\begin{equation}\label{eq:rembd}
\left(\mathcal{G}^{i}\right)_{j}=G\left(s\left(r_{ji}\right),T_{i}, T_{j}\right),
\end{equation}
% \WH{have you tested if the model has the same capacity and is more efficient is we do not input $T_i$? }
where $G$ is a neural network mapping from scalar weight $s\left(r_{j i}\right)$ and type embeddings of both center and neighbor atoms, through multiple hidden layers, to $M_1$ outputs. 
Here we simply feed the concatenated inputs into $G$ at once, as shown in Fig. \ref{fig:model}(b).

\subsection{Attention method for building up trainable descriptors}
\label{subsec:descriptor}
The attention mechanism has achieved great success and played an increasingly important role in CV~\cite{guo2022attention} and NLP~\cite{galassi2020attention}. 
It has become an excellent tool for modeling the importance or relevance of visual regions or text tokens, thus is potentially appropriate to reweight the interactions among neighbor atoms according to both distance and angular information.

In DPA-1, we follow the standard self-attention mechanism and obtain the queries $\mathcal{Q}^{i,l} \in \mathbb{R}^{N_{i}\times d_k}$, keys $\mathcal{K}^{i,l} \in \mathbb{R}^{N_{i}\times d_k}$, and values $\mathcal{V}^{i,l} \in \mathbb{R}^{N_{i}\times d_v}$:
\begin{eqnarray}
\left(\mathcal{Q}^{i,l}\right)_{j}=Q_{l}\left(\left(\mathcal{G}^{i,l-1}\right)_{j}\right),\nonumber\\ \left(\mathcal{K}^{i,l}\right)_{j}=K_{l}\left(\left(\mathcal{G}^{i,l-1}\right)_{j}\right),\\
\left(\mathcal{V}^{i,l}\right)_{j}=V_{l}\left(\left(\mathcal{G}^{i,l-1}\right)_{j}\right),\nonumber
\end{eqnarray}
where $Q_{l}, K_{l}, V_{l}$ represent three linear transformations which output the queries and keys of dimension $d_k$ and values of dimension $d_v$, and $l$ is the index of attention layer. Here we take $\mathcal{G}^{i,0} = \mathcal{G}^{i}$. 

Then we adopt the scaled dot-product attention method\cite{vaswani2017attention} to mix the neighbor features after calculating the attention weights:
\begin{equation}
A(\mathcal{Q}^{i,l}, \mathcal{K}^{i,l}, \mathcal{V}^{i,l}, \mathcal{R}^{i,l})=\varphi\left(\mathcal{Q}^{i,l}, \mathcal{K}^{i,l},\mathcal{R}^{i,l}\right)\mathcal{V}^{i,l},
\end{equation}
where $\varphi\left(\mathcal{Q}^{i,l}, \mathcal{K}^{i,l},\mathcal{R}^{i,l}\right) \in \mathbb{R}^{N_{i}\times N_{i}}$ is attention weights.
In the original attention method, one typically has $\varphi\left(\mathcal{Q}^{i,l}, \mathcal{K}^{i,l}\right)=\operatorname{softmax}\left(\frac{\mathcal{Q}^{i,l} (\mathcal{K}^{i,l})^{T}}{\sqrt{d_{k}}}\right)$, with $\sqrt{d_{k}}$ being the normalization temperature.
This is slightly modified to better incorporate the angular information:
%Here we propose an attention weights gated by angular information $\hat{\mathcal{R}}^{i}(\hat{\mathcal{R}}^{i})^{T} \in \mathbb{R}^{N_{i}\times N_{i}}$:
\begin{equation}
\label{eq:atten_weights}
\varphi\left(\mathcal{Q}^{i,l}, \mathcal{K}^{i,l},\mathcal{R}^{i,l}\right) = \operatorname{softmax}\left(\frac{\mathcal{Q}^{i,l} (\mathcal{K}^{i,l})^{T}}{\sqrt{d_{k}}}\right) \odot \hat{\mathcal{R}}^{i}(\hat{\mathcal{R}}^{i})^{T},
\end{equation}
where $\hat{\mathcal{R}}^{i} = \frac{\mathcal{R}^{i}}{\Vert\mathcal{R}^{i}\Vert_2} \in \mathbb{R}^{N_{i}\times 3}$ denotes normalized relative coordinates and $\odot$ means element-wise multiplication. 
Intuitively, in the neighborhood of center atom $i$, neighbor atom $k$ may be highly correlated with $j$ when both the relative distance attention $(\mathcal{Q}^{i,l})_{j}(\mathcal{K}^{i,l})_{k}^{T}$ and normalized product of relative coordinates $\frac{\textbf{r}_{ji}(\textbf{r}_{ki})^{T}}{r_{ji}r_{ki}}$ have high scores. 

Then we add layer normalization in a residual way to finally obtain the self-attentioned local embedding matrix $\hat{\mathcal{G}}^{i}$ in one such attention layer:
\begin{equation}
\label{eq:attention}
\mathcal{G}^{i,l} = \mathcal{G}^{i,l-1} + \mathrm{LayerNorm}(A(\mathcal{Q}^{i,l}, \mathcal{K}^{i,l}, \mathcal{V}^{i,l}, \mathcal{R}^{i,l})).
\end{equation}

We also tried other attention related tricks such as pre-layer normalization, multi-head attention, etc., which brought little improvement. 
In practice, %we usually repeat $n(n \geq 2)$ times setting as number of attention layers to 
as shown in Fig.~\ref{fig:model}(c), we repeated this procedure by $l(l \geq 2)$ times for a more complete representation. 
If not  stated otherwise, we use $l=2$  in the following sections of the work.
% If no additional instructions are given, we use $l=2$ in the following section. 
%and finally obtain $\hat{\mathcal{G}}^{i} = \mathcal{G}^{i,n}$.

%\subsection{Local descriptor and fitting net}
%\label{subsec:descriptor}
% \begin{figure}
%     \centering
%     \includegraphics[width=0.8\textwidth]{dummy.pdf}
%     \caption{An illustration of our attention-based local descriptor}
%     \label{fig:network_architecture}
% \end{figure}
%As mentioned in \ref{subsec:symmetry}, the original input $\textbf{r}$ should be mapped to symmetry preserving components under translation, rotation and permutation. In original DeePMD, translational and rotational symmetries are preserved in the sub-structure $\tilde{\mathcal{R}}^{i}\left(\tilde{\mathcal{R}}^{i}\right)^{T}$, with relative coordinates and the inner elimination of rotational matrix, respectively. Refer to theorem 2 in \cite{}, the other sub-structure $\left(\mathcal{G}^{i}\right)^{T} \tilde{\mathcal{R}}^{i}$ in DeePMD is a special realization of the permutation invariant operations. Then, similar to DeePMD, 
Next, we define the encoded feature matrix $\mathcal{D}^{i} \in \mathbb{R}^{M_{1} \times M_{2}}$ of atom $i$:
\begin{equation}
\mathcal{D}^{i}=\left(\hat{\mathcal{G}}^{i}\right)^{T} \tilde{\mathcal{R}}^{i}\left(\tilde{\mathcal{R}}^{i}\right)^{T} \dot{\mathcal{G}}^{i},
\end{equation}
where $\dot{\mathcal{G}}^{i}$ stands for a sub-matrix of $\hat{\mathcal{G}}^{i}$, which takes the first $M_2 (< M_1)$ columns of $\hat{\mathcal{G}}^{i}$. Here the feature matrix $\mathcal{D}^{i}$, i.e. the descriptor, preserves all the invariance mentioned above, of which the proof can be found in Ref.~\cite{zhang2018end}. 
We then pass the reshaped $\mathcal{D}^{i}$, concatenated with the type embedding parameters of the center atom, through the multi-layer fitting network:
\begin{equation}
e_{i} = e\left(\mathcal{D}^{i},  T_{i}\right).
\end{equation}
%\LZ{Here we use the symbol $F$ or $e$? We used $e$ in the Formulation Section.}
%where $e_{i}$ is the atomic energy, 
The total energy of the system is then given as the summation of $e_{i}$, and the atomic force $\mathcal{F}_{i}$ can be further computed via Eq.\ref{eq:force}.

\subsection{Model (pre-)training and finetuning}
For model training or pretraining, we adopted the Adam stochastic gradient descent method~\cite{kingma2014adam} on all the trainable parameters $\boldsymbol{w}$ inside the model to minimize the loss:
\begin{equation}
    \mathcal{L}_{\boldsymbol{w}}(E^{\boldsymbol{w}}, \mathcal{F}^{\boldsymbol{w}}) = \frac{1}{\vert\mathcal{B}\vert}\sum_{t \in \mathcal{B}}
    \left( 
    p_{\epsilon}\left \vert E_{t}-E_{t}^{\boldsymbol{w}}\right\vert^{2}
    + p_{f}\left\vert\mathcal{F}_{t}-\mathcal{F}_{t}^{\boldsymbol{w}}\right\vert^{2} 
    \right).
\end{equation}
Here $\mathcal{B}$ represents a minibatch, $\vert\mathcal{B}\vert$ is the batch size, $t$ denotes the index of the training sample. 
$E^{\boldsymbol{w}}, \mathcal{F}^{\boldsymbol{w}}$ denote the model outputs and $E, \mathcal{F}$ are the corresponding DFT results. We also adopted a scheduler to tune the prefactors $p_{\epsilon}$ and $p_{f}$ during the training process to make a better balance between energy and force labels. 
Virial errors, which are omitted here, can be added to the loss for training if available.
%Unsupervised schemes for pretraining the model are left for future investigations.

To finetune the pretrained model with a new dataset, we first change the energy bias in the last layer of the pretrained model with the new statistical results of the new dataset, and then we fix part of the parameters in the pretrained model and train the remaining.
For the following experiments, we obtained the best performance when only the type embedding parameters are fixed.

\section{Experiments}
\label{sec:experiments}
We conducted a number of experiments to evaluate the performance of DPA-1.
First, to test the model's ability to transfer among different compositions, we trained it from scratch against various systems and tested it under several challenging schemes.
Then, we used an AlMgCu dataset to test its ability to transfer to ternary systems upon pretraining with single-element and binary data.
Finally, we pretrained DPA-1 using the OC2M subset in OC20 dataset~\cite{chanussot2021open} and applied it to various downstream tasks.
To illustrate the effectiveness of the type-embedding and attention schemes, we compared against DeepPot-SE model~\cite{zhang2018end} in all the experiments. 
In the following, we shall introduce first the datasets we used, and then the experiments we conducted.
%Next we will list the results on several datasets with different experimental settings, to show the superiority of our proposed model DP-Mendeleev(marked as \textbf{DPM}).
%We deliberately designed tests on three levels of transferability mentioned before, which are much more challenging for other ML-based models such as DeePMD.
%In order to make it detail, we also perform experiments in an active learning style, to show the accuracy especially when the amount of available data is relatively small. 
%Results reveal that our model is able to encode more configuration and chemical information from existing training data, and thus outperforms DeePMD on all the tests on transferability, with improvements even in order of magnitudes under circumstances. 
%With foundations on transferability and capacity of element types, we thus provided a large-scale pretrained model with 56 elements on OC20 dataset and together with a novel transfer learning procedure through which we show the reusability of this pretrained model on datasets containing totally different compositions. 
%Furthermore, a visualization of learned type embedding and interpolation in the corresponding latent space are performed to confirm the interpretability of the pretrained model.

%\WH{The second advantage we want to demonstrate is that the DPM encodes more configurations and chemical information of existing training data in the pre-training stage, thus requires substantially smaller amount of training data in downstream tasks. This information should be explicitly delivered} \ZD{Now?}

\begin{table*}[]
    \caption{Validation RMSE of DPA-1 and DeepPot-SE on energy (meV/atom) and atomic forces (meV/\AA) with different settings of the training/validation sets (See Sec. \ref{dataset_intro} for details). The number of attention layers $l$ in DPA-1 is set to 2 in the AlMgCu and SSE systems, and to 3 in the HEA systems.
    %The results are given by DPA-1 and DeepPot-SE outside and inside the brackets, respectively.
    %Results given by DeepPot-SE are shown in bracket. 
    %The component of subsets are presented in section \ref{dataset_intro}, which are further divided into corresponding training and validation sets, thus we use the name of subsets here only to represent where samples are come from. 
    %Detailed introduction of the datasets can be found in Sec. \ref{dataset_intro}.
    Bold numbers correspond to lower values. 
    %The results reveal the superiority of DPM on the transferability between different subsets of samples.
    %\WH{different columns to present DPM and DP}
    %\WH{THis HEA should be distinguished from the HEA in Tab. 1}
    }
    \label{table_trans}
    \centering
\begin{tabular}{ccccccc}
\hline
\multirow{3}{*}{\textbf{Systems}} & \multirow{3}{*}{\textbf{Training}} & \multirow{3}{*}{\textbf{Validation}} & \multicolumn{4}{c}{\textbf{Validation RMSE}}                                                                                                                                                                                                  \\ \cline{4-7} 
                                  &                                    &                                      & \multicolumn{2}{c}{\textbf{DPA-1}}                                                                                    & \multicolumn{2}{c}{\textbf{DeepPot-SE}}                                                                               \\ \cline{4-7} 
                                  &                                    &                                      & \begin{tabular}[c]{@{}c@{}}Energy\\ (meV/atom)\end{tabular} & \begin{tabular}[c]{@{}c@{}}Force\\ (meV/\AA)\end{tabular} & \begin{tabular}[c]{@{}c@{}}Energy\\ (meV/atom)\end{tabular} & \begin{tabular}[c]{@{}c@{}}Force\\ (meV/\AA)\end{tabular} \\ \hline
\multirow{3}{*}{AlMgCu}           & single + binary                    & ternary                              & \textbf{6.99}                                               & \textbf{58}                                             & 65.1                                                        & 92                                                      \\
                                  & all (single + binary +  ternary)   & ternary                              & \textbf{2.26}                                               & \textbf{35}                                             & 3.16                                                        & 42                                                      \\
                                  & all                                & all                                  & \textbf{2.74}                                               & \textbf{38}                                             & 3.67                                                        & 45                                                      \\ \hline
\multirow{3}{*}{SSE}              & init + single                      & mix                                  & \textbf{0.56}                                               & \textbf{60}                                             & 0.72                                                        & 76                                                      \\
                                  & init + mix                         & single                               & \textbf{3.72}                                               & \textbf{69}                                             & 3.76                                                        & 82                                                      \\
                                  & all (init + single + mix)          & all                                  & \textbf{1.41}                                               & \textbf{68}                                             & 2.92                                                        & 85                                                      \\ \hline
\multirow{3}{*}{HEA}              & interior                           & exterior                             & \textbf{31.2}                                               & \textbf{158}                                             & 197                                                         & 399                                                     \\
                                  & exterior                           & interior                             & \textbf{6.88}                                               & \textbf{117}                                             & 236                                                        & 428                                                     \\
                                  & all (interior + exterior)          & all                                  & \textbf{4.96}                                               & \textbf{71}                                             & 28.7                                                        & 141                                                     \\ \hline
\end{tabular}
\end{table*}
\begin{figure*}
    \includegraphics[width=0.9\textwidth]{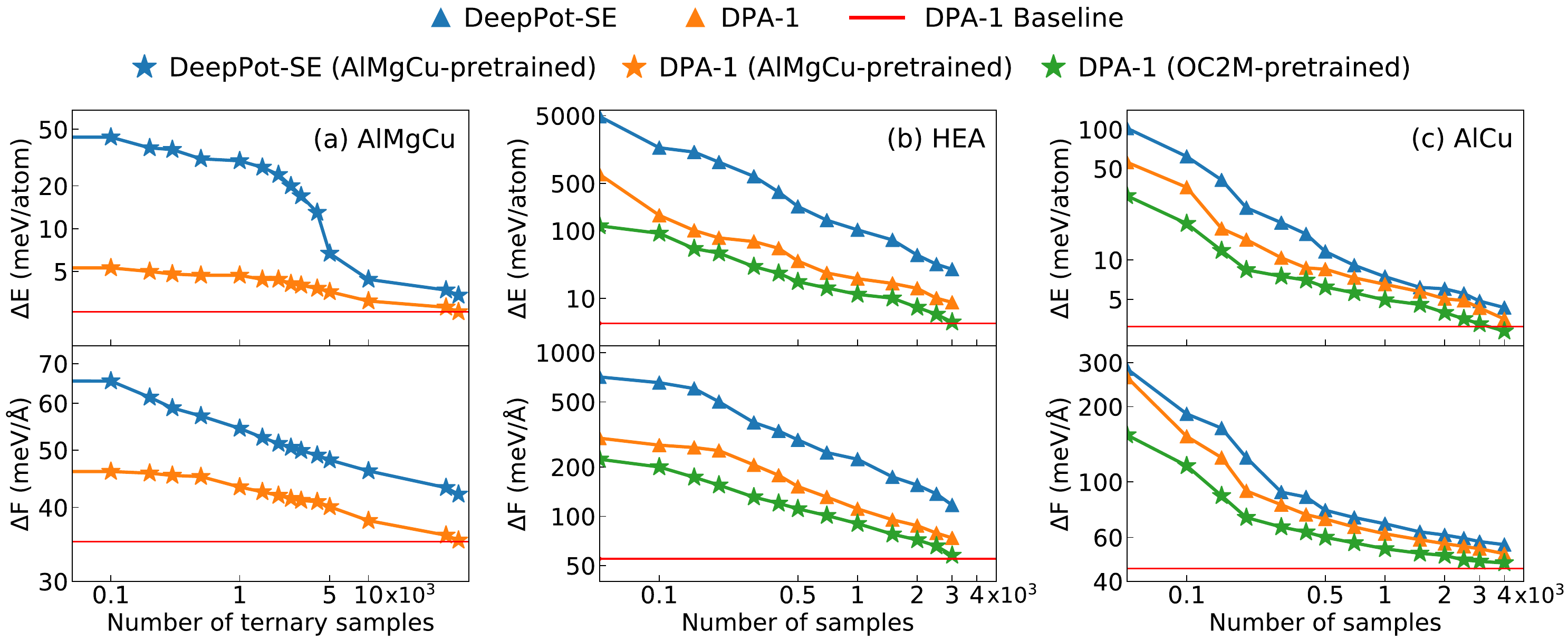}
    \centering
    \caption{
    Learning curves of both energy and force
    with DeepPot-SE and DPA-1, under different setups and on different systems.
    (a) Learning curves on the AlMgCu ternary subset, with DeepPot-SE and DPA-1 models pretrained on single-element and binary subsets;
    (b-c) Learning curves on HEA (b) and AlCu (c), with DeepPot-SE (from scratch) and DPA-1 (both from scratch and pretrained on OC2M). 
    Red line represents the full-data-training baseline with DPA-1. 
    }
    \label{fig:learning-curve}
\end{figure*}
\begin{figure*}
    \includegraphics[width=0.8\textwidth]{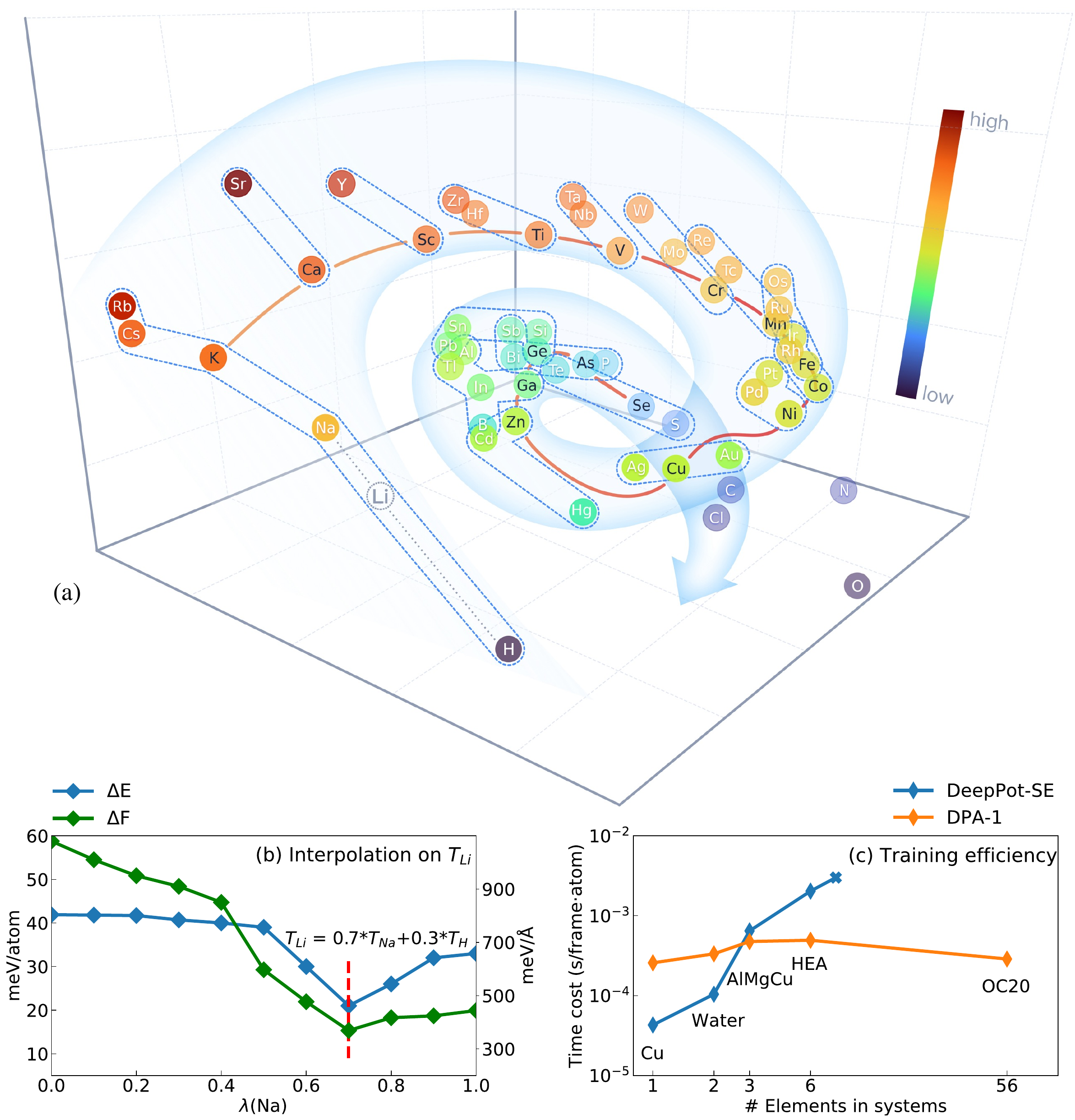}
    \centering
    \caption{
    (a) 3-dimensional PCA visualization of the learned type embeddings of DPA-1 pretrained on OC2M. 
    These 56 elements are roughly arranged on a \textbf{\emph{spiral}} in the latent space. 
    Elements in the fourth period are connected with the red line and elements belonging to the same family are grouped by the blue dot lines. 
    Colors on the names of the elements represent the height in $z$-axis. 
    We use dashed circle to denote the hypothetical position of Li, which is not contained in OC2M.
    See text for discussions.
    (b) RMSE of energy and force for SSE systems given by DPA-1 pretrained on OC2M, as functions of linear interpolation coefficient $\lambda\left(Na\right)$.
    Since Li is not contained in OC2M, we let $T_{Li} = \lambda\left(Na\right)*T_{Na} +\left(1-\lambda\left(Na\right)\right)*T_{H}$ be the interpolated type embedding of Li.
    The OC2M-pretrained model with this interpolation and modified energy bias is directly tested on SSE systems without further training. 
    (c) Training efficiency of DPA-1 and DeepPot-SE (considering type information of both two sides) with the growing number of element types in training systems. 
    The maximum number of neighboring atoms to be considered is set to 120 in all the experiments. 
    %DeepPot-SE is more efficient than DPA-1 at the beginning and grows rapidly with the number of element types for the $O(N^2)$ complexity, while the efficiency of DPA-1 is nearly irrelevant to the element types and thus can model dozens of elements on the periodic table with ease.
    }
    \label{fig:interpretability}
\end{figure*}

\subsection{Datasets}
\label{dataset_intro}
%We first introduce all the datasets to perform our experiments on, detailed information of which can be found in the corresponding references. 
%Note that these systems are much more easier than the other following datasets, considering the complexity of both configurations and compositions.

% \textbf{AlMgCu alloy systems.} This dataset is generated using DP-GEN, a concurrent learning scheme, after exploring 2.73 billion alloy configurations, only a small portion ($\sim$100k configurations for $\sim$2000 bulk and surface systems) of which are labeled and contained. The exploration runs in the whole concentration space, i.e., \ce{Al_xCu_yMg_z} with $0 \leq x,y, z \leq 1,x+y+z=1$, and the configuration space covers a temperature range around 50.0 K to 2579.8 K and a pressure range around 1 bar to 50000 bar. The computational cost during data generation is vastly high, including the exploring, training and labeling procedures running on both GPUs and CPUs for months. 
% % This eagerly calls for a more efficient sampling and learning method through models with better generalization ability, being the original purpose of this proposed method. 
% In section \ref{active}, we show an exciting trend in active learning procedure that our model can hopefully reduce the amount of necessary data to lower than 10\%.

\textbf{AlMgCu alloy systems~\cite{jiang2021accurate}.} This dataset is generated using DP-GEN~\cite{zhang2020dp}, a concurrent learning scheme. After exploring 2.73 billion alloy configurations (derived from $\sim$2000 bulk and surface systems), only a small portion ($\sim$100k configurations ) of them are labeled then compose the compact dataset.
%, after exploring 2.73 billion alloy configurations (derived from $\sim$2000 bulk and surface systems), only a small portion ($\sim$100k configurations ) of which are labeled and contained. 
The exploration runs in the whole concentration space, i.e., \ce{Al_xCu_yMg_z} with $0 \leq x,y,z \leq 1,x+y+z=1$, and $x$, $y$, $z$ take discrete values permitted by the finite-size simulation boxes. We can divide the systems into \textbf{\emph{single}}, \textbf{\emph{binary}} and \textbf{\emph{ternary}} subsets, in name of the number of non-zero $x$, $y$, $z$.
The configuration space covers a temperature range around 50.0 K to 2579.8 K and a pressure range around 1 bar to 50000 bar. 
%The computational cost during data generation is vastly high. The workflow including the exploring, training and labeling procedures comsumes over thounsands of GPU card hours and over ten million CPU core hours, which takes months using both the resources of the high-performance computing cluster and the high-throughput cloud computing platform. 
%In section \ref{active}, we show an exciting trend in active learning procedure that our model can hopefully reduce the amount of necessary data to lower than 10\%.

\textbf{Solid-state electrolyte (SSE) systems~\cite{huang2021deep}.} These systems contain \ce{Li10XP2S12}-type SSE materials, where \ce{X} represents single or combination of \ce{Ge}/\ce{Si}/\ce{Sn}, and can be divided into three main parts: \textbf{\emph{init}}, \textbf{\emph{mix}} and \textbf{\emph{single}}.
The \emph{init} part comes from standard DP-GEN scheme starting from 590 structures that are generated via slightly perturbing DFT-relaxed crystal structures \ce{Li10Ge(PS6)2}, \ce{Li10SiP2S12} and \ce{Li10SnP2S12} from Materials Project~\cite{jain2013commentary}. The exploration covers both ordered structures relaxed by DFT (i.e. structures downloaded from the Materials Project database, in which the position of \ce{Ge}/\ce{Si}/\ce{Sn}/\ce{P} atoms are fixed) and disordered structures whose 4d sites are randomly occupied by \ce{Ge}/\ce{Si}/\ce{Sn}/\ce{P}. Based on the \emph{init} part, the \emph{mix} part contains further exploration in binary and ternary mixture of \ce{Ge}/\ce{Si}/\ce{Sn}, while the \emph{single} part covers only single X in \ce{Ge}/\ce{Si}/\ce{Sn} with other changes in lattice and ratio of \ce{Li}. 

\textbf{HEA systems.} 
%Similar to SSE systems above, HEA systems include DP-GEN exploration of both configurations and compositions of bulk \ce{TaNbWMoVAl} alloy, mainly divided into two groups: \textbf{\emph{interior}} and \textbf{\emph{exterior}}. \ce{Ta_3Nb_3W_3Mo_3V_3Al_1} is set as the central point of the composition space, given a 16-atom simulation box containing the former 5 elements as main components and \ce{Al} as the additive. The \emph{interior} group includes the central point and nearby composition variations, covering six-component, quinary, quaternary and ternary alloys with relatively mild 
%concentrations of each component. The \emph{exterior} group includes more distant composition variations where one or two elements dominate, and reaches the edges and corners of the composition space as binary alloys and simple substances of main components. For both groups, the temperature range is around 50.0 K to 388.1 K and the pressure range is around 1 bar to 50000 bar.
% a revision of the HEA section
The High Entropy Alloy HEA dataset includes bulk \ce{TaNbWMoVAl} alloy systems of various configurations and compositions. 
We employ DP-GEN to explore the composition space, starting from \ce{Ta_3Nb_3W_3Mo_3V_3Al_1}, a 16-atom unit cell containing the former 5 elements as main components and \ce{Al} as an additive. 
The dataset is divided into two subsets: \textbf{\emph{interior}} and \textbf{\emph{exterior}}. 
The \emph{interior} (higher entropy) subset includes composition variations near the starting point. 
It covers six-component, quinary, quaternary and ternary alloys.
The \emph{exterior} (lower entropy) subset includes systems that are close to the corners and edges of the composition space. 
It includes systems where one or two elements dominate, binary alloys and simple substance systems.
 For both subsets, the temperature range is around 50.0 K to 388.1 K and the pressure range is around 1 bar to 50000 bar.

%We particularly focus on the transferability of the model between these two groups, namely two different parts of the composition space.
%\WH{mention that the datasets are generated by DP-GEN scheme.}

%\textbf{Open Quantum Materials Database (OQMD)~\cite{saal2013materials}.} OQMD is a database of DFT calculated thermodynamic and structural properties of 1,022,603 materials with an amount of 89 elements. This database can be a challenge for model capacity of atomic species, while it fails to be a proper dataset for learning inter-atomic energies and forces as it only contains systems at equilibrium, where all forces are zero. We merely show the training results on energies.

\textbf{OC20~\cite{chanussot2021open}.} OC20 consists of single adsorbates (small molecules) physically binding to the surfaces of catalysts covering periodic bulk materials with 56 elements. Both the chemical diversity and system size are much more complex than other benchmark datasets, such as MD17~\cite{chmiela2017machine}, ANI-1x~\cite{smith2018less} or QM9~\cite{ramakrishnan2014quantum}. 
OC2M is a subset including 2 million data points (energies and forces) randomly sampled from OC20, which is still challenging for model training and decent for pretraining. 
Johannes et al. recently provided several baselines on OC2M, taking months to converge\cite{gasteiger2022graph}. 
%\LZ{give ref? and also refs for MD17, ANI-1x or QM7?}
%We show in section \ref{pretrain} that our model achieves comparable results after days of training and we also propose a universal pretraining procedure to reuse the trained model as a start point for brand new tasks. 
%Active learning on both \ce{AlCu} and \ce{AlMgCu} systems from the beginning shows the effect of pretraining, which may be a paradigm of modeling inter-atomic potential in a transfer learning style. 

\subsection{Accuracy on various datasets, trained from scratch}
%, which confirms the superiority to transfer among different configurations.
%The first column of the numbers represents the finetuned results from the pretrained model of DPA-1 on OC20 dataset, details of which will be discussed in section \ref{pretrain}. 

% \WH{Again, ``transferability". I would not suggest you using it. 
% In the force field community, the transferability is basically a mixture of in-distribution generalizability and out-of-distribution generalizability. 
% When you say it they will understand in a totally different way.
% In this work, we do not want to stress on the generalizability of the DPM model, but its ability of learning knowledge (configurational and chemical) from the data set, and achieve better performance in downstream tasks.
% }
The majority of existing models usually focus on the ability to transfer among different configurations, in which case training and validation subsets consist of similar compositions (e.g. randomly sampled from the same dataset).
However, to perform pretraining, the upstream and downstream datasets may differ violently.
Thus, it's vital for models under pretraining scheme to transfer among different compositions or even among different datasets, which has, as far as we know, rarely been discussed before.
In this work, we mainly focus on a more general but challenging scheme to comprehensively test the generalization ability of the model. 

We first designed several challenging tasks to test the model's ability to transfer among different compositions. 
For AlMgCu, SSE, and HEA systems, we divided them into subsets with different compositions for training and validation (See Sec. \ref{dataset_intro} for details).
The results of DPA-1 and DeepPot-SE are shown in Table \ref{table_trans}.
%Note that these challenging tests require a more general understanding and embedding of inter-atomic interactions, in some of which DeepPot-SE has failed. 
With the training loss nearly the same (omitted in the table), the DPA-1 drastically outperforms DeepPot-SE in the validation accuracy. 
For example, for AlMgCu systems, when trained only on single- and binary-element samples, the validation RMSE of DPA-1 on ternary samples can outperform DeepPot-SE by one order of magnitude (6.99 versus 65.1 meV/atom). 
%the validation RMSE of DPA-1 and DeepPot-SE may differ in orders, for example, 6.99 vs. 65.1 meV/atom prediction RMSE on ternary samples of AlMgCu systems using models trained only on single-element and binary samples. 
This suggests that the DPA-1 model might have learned the latent interactions of ternary pairs Al-Mg-Cu from binary pairs Al-Mg, Al-Cu, Mg-Cu, and single-element interactions, possibly thanks to the type-embedding scheme and attention mechanism.   \recheck{We conducted an ablation study in Appendix.~\ref{Asec:ablation} on HEA systems to demonstrate the influence of each structural component.}
% For HEA systems, we use more attention layers ($l=3$) due to their complexity, where much more outperformance can be seen (6.88 versus 236 meV/atom) under some specific settings.
% and a more general embedding of element types.
%The situations are similar in SSE and HEA systems, showing the superiority of DPA-1 to transfer between different subsets of samples, which is key to reduce the amount of necessary data samples for training a complete and accurate potential.
%Besides, we can also see the improvements in full-data-training scenes in the last row of each system in Table \ref{table_trans}, showing the universality of the proposed model structure. 

To test the performance of DPA-1 in terms of predicting more physical quantities, 
\recheck{we performed geometry relaxations on all AlMgCu ternary alloys available from the Materials Project to evaluate their accuracy in predicting formation energy and equilibrium volume (see details in Appendix~\ref{Asec:relaxations}).}
\recheck{We also} used it to calculate the elastic moduli of AlMgCu systems, which requires accurately capturing the second-order information (see details in Appendix~\ref{Asec:elastic moduli}). 
  \recheck{
Additionally, we carried out molecular dynamics simulations on LiGePS systems to assess the diffusion coefficients in relation to temperature, comparing the results to \textit{ab initio} molecular dynamics (AIMD) simulations and experimental studies (see details in Appendix~\ref{Asec:diffusion coefficients}).
In all tests, satisfactory agreement with the DFT and/or experimental references are obtained.} 

As a supplement, we also trained DPA-1 model on several simple systems to compare with other ML-based PES. Since these tasks are much easier than the above ones and out of our main focus, we place the results in Appendix Section~\ref{Asec:simple datasets}. Note that there may be relatively little room for improvement on these simple datasets, while DPA-1 can still outperform other methods with even less training samples. 

\subsection{Sample efficiency of pretrained models}
%\WH{This title is misleading. I would use something like ``downstream tasks"}
\label{subsec:sample efficiency}
As shown in Fig.~\ref{fig:learning-curve}, we use the learning curves to illustrate in terms of the amount of additional training data saved for downstream tasks thanks to model pretraining.
In all the experiments, the learning curves were generated by an active learning procedure, in which a pool of data labeled by energy and force is prepared and three steps are repeated iteratively: 
using samples in the training pool to train the model; 
testing the model using the remaining samples;
selecting 50 samples with the largest prediction errors on per-atom energies and adding them to the training pool.
We use the term sample efficiency to denote the amount of training samples required by a model to achieve a given accuracy level for a certain task.

We started with a relatively simple task to compare DeepPot-SE and DPA-1.
In this task, both the two models were pretrained using 
single-element and binary subsets of the AlMgCu systems, and the learning curves were obtained using the AlMgCu ternary subset.
As shown in Fig.~\ref{fig:learning-curve}(a), DPA-1 exhibits a much better sample efficiency than DeepPot-SE, which should be expected.

Next, we used the OC2M dataset, which contains 56 elements, to pretrain DPA-1 and evaluated its performance on the HEA systems and the AlCu systems (Figs.~\ref{fig:learning-curve}(b) and (c), respectively).
As shown in Fig.~\ref{fig:interpretability}(c), the training cost of DeepPot-SE scales quadratically with the number of elements, making its pretraining computationally infeasible, while the number of elements has no effects on the training cost of DPA-1. 
It is observed that the sample efficiency of DPA-1 pretrained on OC2M is generally better than DPA-1 from scratch, while DeepPot-SE from scratch is the worst.
Moreover, compared with AlCu systems, the improvement of pretraining is much more significant for HEA systems, possibly due to the fact that the number of elements of HEA is much larger than AlCu, and the local chemical environment is much more complicated.

\recheck{The equivariant GNN models usually needs thousands of GPU hours to be trained to a descent accuracy~\cite{gasteiger2022graph}. 
By contrast, the DPA-1 modle only takes less than 200 GPU hours for training.
The converged energy and force MAEs on the OC2M validation set are 0.681~eV and 0.076 eV/\AA, respectively. 
This accuracy is  comparable with the best energy conserving GNN model DimeNet++, which achieves MAEs of 0.805~eV and 0.066 eV/\AA, reported in Ref.~\cite{gasteiger2022graph}.
A better performance of energy MAE 0.286~eV and force MAE 0.026 eV/\AA\ is achieved by GemNet-OC at the cost of non-conservative forces and loss of smoothness~\cite{gasteiger2022graph}. 
}

\subsection{Interpretability of type embedding learned from pretraining}
\label{subsec:interpretability}
To see whether DPA-1 can learn physically meaningful information from pretraining, we investigated the 3-dimensional principal component analysis (PCA) visualization of the learned type embeddings in the OC2M-pretrained model.
Interestingly, as shown in Fig.~\ref{fig:interpretability}(a), the arrangement of the elements generally follows the shape of a downward spiral.
Elements belonging to the same period are lined up in the direction of the spiral;
while elements belonging to the same family are listed in the direction orthogonal to the spiral.
%shows the 3-dimensional principal component analysis (PCA) visualization of the learned type embeddings in the OC2M-pretrained model.
%aiming to show the interpretability towards the theoretical distribution of periodic table. 
%Surprisingly, 
%the outer electrons of each element horizontally grow, and in the meantime, vertically equal along the spiral, which amazingly correspond to the periods and families of the period table of elements, respectively. 
Even though some transition metal elements are almost bounded together, this rule still roughly holds.
%due to the similar environments in their neighborhoods, yet still can be grouped with nearest elements in the fourth period. 
It is observed that C, N and O are outliers, possibly because in OC2M, C, N and O are mostly in organic molecules, which serve as adsorbates and have chemical environments that are very different from other elements.
%only occupy a small part of OC2M (as adsorbates). 

In addition, 
%observing the distribution of learned type embeddings in the pretrained model, 
we performed interpolation experiments for the type embedding of Li, an element unseen in OC2M.
As shown in Fig. \ref{fig:interpretability}(b), we let $T_{Li} = \lambda\left(Na\right)*T_{Na} +\left(1-\lambda\left(Na\right)\right)*T_{H}$, since Li lies between H and Na in the same family.
When tested on the SSE system, only the bias in the atomic energy is changed, since the setup of the electronic method used to label the SSE system is different from that for OC2M, which typically causes an energy shift.
%unseen elements in the latent space and tested the accuracy after changing the energy bias in transfer learning yet without further training. 
%Fig. \ref{fig:interpretability}(b) shows the interpolation of Li, excluded in OC2M, with $T_{Li} = \lambda\left(Na\right)*T_{Na} +\left(1-\lambda\left(Na\right)\right)*T_{H}$ for the sake of congeners. 
It is found that the RMSE of energy and force shows a sudden drop when $\lambda\left(Na\right) = 0.7$, which meets the chemical intuition and further confirms the interpretability of the pretrained DPA-1 model. 
  \recheck{Moreover, we conducted analogous interpolation experiments for Nb and Mo on the HEA systems, and reached similar conclusions as the Li interpolation (see detailed report  in Appendix.~\ref{Asec:interpolation}).
}

\label{pretrain}

\section{Summary}
In this paper, we developed DPA-1, an attention-based Deep Potential model that allows for large-scale pretraining on atomistic datasets.
We tested DPA-1 from different aspects, showing its excellent performance in terms of its accuracy on various datasets when trained from scratch, as well as its sample efficiency when pretrained with existing data.
Further investigations on the type embedding parameters suggests the interpretability of DPA-1 pretrained on OC2M.

In the future, it will be of interest to extend the training dataset to cover the full periodic table, and, in particular, see a more converged ``spiral'' in the latent space;
the embedding information of local chemical environments may be useful to characterize different conformations. 
Multi-task and unsupervised training schemes would worth exploring; 
and, for downstream tasks, just like what has happened in the fields of CV and NLP, schemes like model compression, distillation, and transfer, etc., are desperately needed.
We leave these possibilities and more applications to future works.

\section{Data availability}
The dataset used for training OC2M-pretrained DPA-1 is available at: ~\url{https://www.aissquare.com/datasets/detail?pageType=datasets&name=Open_Catalyst_2020(OC20_Dataset)}. Other datasets are available in their references or on reasonable request.

\section{Code availability}
The codes of DPA-1 are in the repository of DeePMD-kit:~\url{https://github.com/deepmodeling/deepmd-kit}. The OC2M-pretrained model is available at:~\url{https://www.aissquare.com/models/detail?pageType=models&name=DPA_1_OC2M}. All the experiments are performed on single Nvidia Tesla V100.

\section{AUTHOR CONTRIBUTIONS}
D.Z, L.Z, H.W. and F.Z.D. conceived the idea of this work. D.Z., H.B. and H.W. designed the model structure. D.Z. implemented the model. D.Z., H.B. and W.J performed the experiments on different systems. All authors contributed to the discussions and edited the manuscript.

\section{COMPETING INTERESTS}
The authors declare no competing financial or non-financial interests.

\begin{acknowledgments}
The work of H.W.~was supported by the National Science Foundation of China under Grant No.11871110 and 12122103. 
We thank Y.L., Z.L. and G.K. for inspiring discussions.
The computational resource was supported by the Bohrium Cloud Platform at DP technology.
\end{acknowledgments}

% \bibliography{ref}{}
% \bibliographystyle{unsrt}

\clearpage
\appendix
\setcounter{table}{0}
\setcounter{figure}{0}
\renewcommand\thefigure{A.\arabic{figure}}
\renewcommand\thetable{A.\arabic{table}}

\section{  \recheck{Ablation study}}
\label{Asec:ablation}

\begin{table*}[]
    \caption{  \recheck{
    Ablation study of the DPA-1 model architecture. The models are validated by the energy (meV/atom) and atomic forces (meV/\AA) RMSEs calculated with different settings of the training/validation sets.
    Bold numbers correspond to lowest values. 
    }
    }
    \label{table_ablation}
    \centering
\begin{tabular}{ccccccc}
\hline
\textbf{}                                                                       & \multicolumn{6}{c}{\textbf{HEA validation RMSE}}                                                                                                                                                                                                                                                                                                                      \\ \hline
Training                                                                        & \multicolumn{2}{c}{\textbf{interior}}                                                                                 & \multicolumn{2}{c}{\textbf{exterior}}                                                                                 & \multicolumn{2}{c}{\textbf{all}}                                                                                      \\ \hline
Validation                                                                      & \multicolumn{2}{c}{\textbf{exterior}}                                                                                 & \multicolumn{2}{c}{\textbf{interior}}                                                                                 & \multicolumn{2}{c}{\textbf{all}}                                                                                      \\ \hline
                                                                                & \begin{tabular}[c]{@{}c@{}}Energy\\ (meV/atom)\end{tabular} & \begin{tabular}[c]{@{}c@{}}Force\\ (meV/Å)\end{tabular} & \begin{tabular}[c]{@{}c@{}}Energy\\ (meV/atom)\end{tabular} & \begin{tabular}[c]{@{}c@{}}Force\\ (meV/Å)\end{tabular} & \begin{tabular}[c]{@{}c@{}}Energy\\ (meV/atom)\end{tabular} & \begin{tabular}[c]{@{}c@{}}Force\\ (meV/Å)\end{tabular} \\ \hline
DeepPotSE                                                                       & 141.32                                                      & 257                                                     & 40.74                                                       & 204                                                     & 15.75                                                       & 117                                                     \\ \hline
\begin{tabular}[c]{@{}c@{}}DPA-1 w/o attention\\ (\texttt{tebd})\end{tabular}            & 88.07                                                       & 252                                                     & 23.66                                                       & 157                                                     & 7.69                                                        & 88                                                      \\ \hline
\begin{tabular}[c]{@{}c@{}}DPA-1 w/o gated\\ (\texttt{tebd}+\texttt{attention})\end{tabular}      & 34.79                                                       & 191                                                     & 9.59                                                        & 120                                                     & 6.35                                                        & 75                                                      \\ \hline
\begin{tabular}[c]{@{}c@{}}DPA-1 gaussian\\ (\texttt{tebd}+\texttt{attention}+\texttt{gated})\end{tabular} & 35.55                                                       & 161                                                     & 7.01                                                        & 121                                                     & 4.97                                                        & 77                                                      \\ \hline
\begin{tabular}[c]{@{}c@{}}DPA-1 full\\ (\texttt{tebd}+\texttt{attention}+\texttt{gated})\end{tabular}     & \textbf{31.27}                                              & \textbf{158}                                            & \textbf{6.88}                                               & \textbf{117}                                            & \textbf{4.96}                                               & \textbf{71}                                             \\ \hline
\end{tabular}
\end{table*}

  \recheck{
We conducted an ablation study on the model structure of DPA-1 for HEA systems, as illustrated in Table~\ref{table_ablation}. 
Our focus was on three primary components that make the DPA-1 different from the DeepPot-SE model: type embedding defined by Eq.~\ref{eq:tebd} (denoted as \texttt{tebd}), attention updates by Eq.~\ref{eq:attention} (denoted as \texttt{attention}), and the gate of $\mathrm{\hat{\mathcal{R}}^{i}(\hat{\mathcal{R}}^{i})^{T}}$ in Eq.~\ref{eq:atten_weights} (denoted as \texttt{gate}).
The full DPA-1 structure is thus ``\texttt{tebd}+\texttt{attention}+\texttt{gate}".
Our findings indicate that \texttt{tebd} and \texttt{attention} substantially enhance performance, potentially due to the effective encoding of elemental space and the communication of information within the neighborhood. 
The \texttt{gate} component in the attention map encodes the angular information of neighboring atoms. 
It provides a relatively smaller but not negligible contribution to the model accuracy. 
Moreover, we examined the impact of different encodings of the radial information by
altering the neural network defined embedding (i.e.~Eq.~\eqref{eq:rembd}) to a Gaussian kernels defined embedding (denoted by DPA-1 gaussian). 
Note that the distance between atoms are not inversed and scaled as Eq.~\ref{eq:swf}, but affine transformed with element-type-dependent parameters~\cite{zhou2022uni}. 
% solely altering the switch function in Eq.~\ref{eq:swf} to Gaussian kernels, denoted as \textit{DPA-1 gaussian}. 
A slight but observable reduction in the test accuracy is observed.
This change renders the model non-smooth under the periodic boundary condition, i.e. a sudden energy jump may occur as the positions of atoms vary infinitesimally. 
This leads to a non-conserved energy in the Hamiltonian dynamics simulations.
% ，which is used in computing the dynamical properties like diffusion constant and viscosity.
}

\section{  \recheck{Equilibrium properties of tenary AlMgCu alloys}}
\label{Asec:relaxations}
\begin{figure*}
    \includegraphics[width=1.0\textwidth]{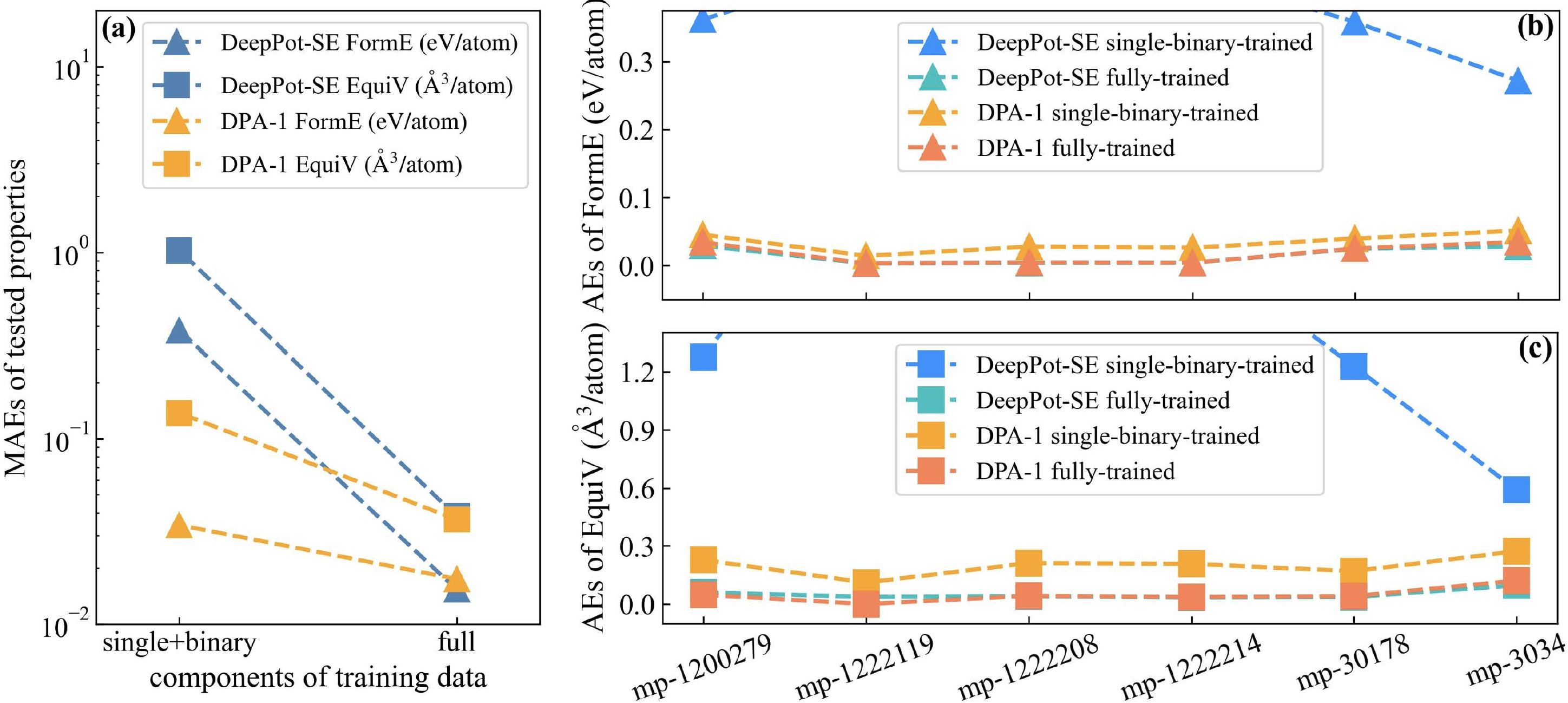}
    \centering
    \caption{  \recheck{(a) Mean absolute errors (MAEs) of formation energy (FormE) and equilibrium volume (EquiV) for all AlMgCu ternary alloys from the Materials Project, as predicted by single-binary-trained and fully-trained DPA-1 and DeepPot-SE models after relaxation. (b) (c) Structure-resolved absolute errors (AEs).}}
    \label{fig:relax_AlMgCu}
\end{figure*}

  \recheck{
We performed geometry relaxation calculations on 
%six AlMgCu alloys 
all AlMgCu ternary alloys available from the Materials Project~\cite{jain2013commentary}. 
We compared single-binary-trained and fully-trained DPA-1 and DeepPotSE models on the accuracy of formation energy and equilibrium volume. 
Mean absolute errors (MAEs) in Fig~\ref{fig:relax_AlMgCu}~(a) shows that DPA-1 model trained only on single and binary samples, achieves 
%decent accuracy on ternary samples, which is much better than DeepPot-SE, 
remarkable enhancement in accuracy on ternary alloys compared with the single-binary-trained DeepPot-SE, confirming DPA-1's superior generalization capabilities.
The enhancement is around one order of magnitude for both properties, from about 400~meV/atom to less than 40~meV/atom in formation energy and from about 1 \AA$^3$/atom to about 0.1 \AA$^3$/atom in equilibrium volume.
%if detailed figure is also supplied, the following text could serve:
Structure-resolved errors in Fig~\ref{fig:relax_AlMgCu} (b)(c) clearly indicate that, the performance of the DPA-1 model on all six ternary alloys, is comparable with that of the fully trained DPA-1 or DeepPot-SE. 
% As the underneath quantities to evaluate the formation energy, per-atom energy and tested structure of stable single-element bulks are given. 
% Detailed errors of single-binary-trained DeepPot-SE abruptly increase when testing ternary structures, being out of the plotting range for both properties.
}

\section{Test on the elastic moduli of AlMgCu alloys}
\label{Asec:elastic moduli}
\begin{figure*}
    \includegraphics[width=0.8\textwidth]{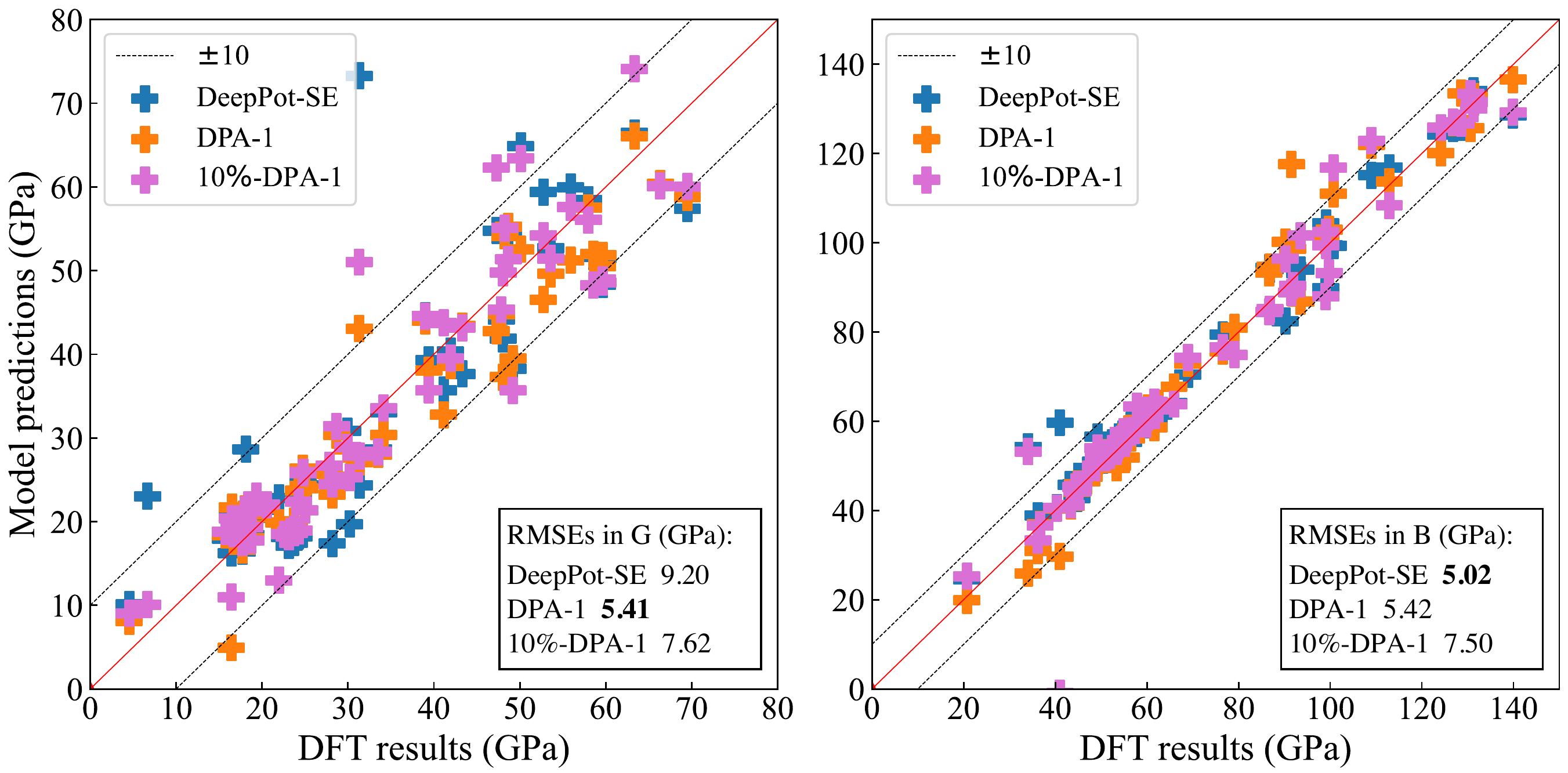}
    \centering
    \caption{Test of trained DeepPot-SE and DPA-1 models on shear (a) and bulk (b) moduli of AlMgCu systems. Both DeepPot-SE and DPA-1 models are trained on all subsets of AlMgCu systems, while 10\%-DPA-1 model used less ternary samples (10\%).}
    \label{fig:elastic}
\end{figure*}

Test results of elastic moduli of \recheck{binary and tenary} AlMgCu \recheck{alloys} are shown in Fig.~\ref{fig:elastic}, with the overall RMSE of both moduli are respectively 7.41, \textbf{5.41} and 7.56 GPa for DeepPot-SE, DPA-1 and 10\%-DPA-1. Finite difference method is used to calculate the elastic moduli, where the maximum norm and shear deformation with respect to the equilibrium conformation are both 2\%.

\section{  \recheck{Diffusion coefficients on LiGePS systems}}
\label{Asec:diffusion coefficients}
\begin{figure*}
    \includegraphics[width=0.46\textwidth]{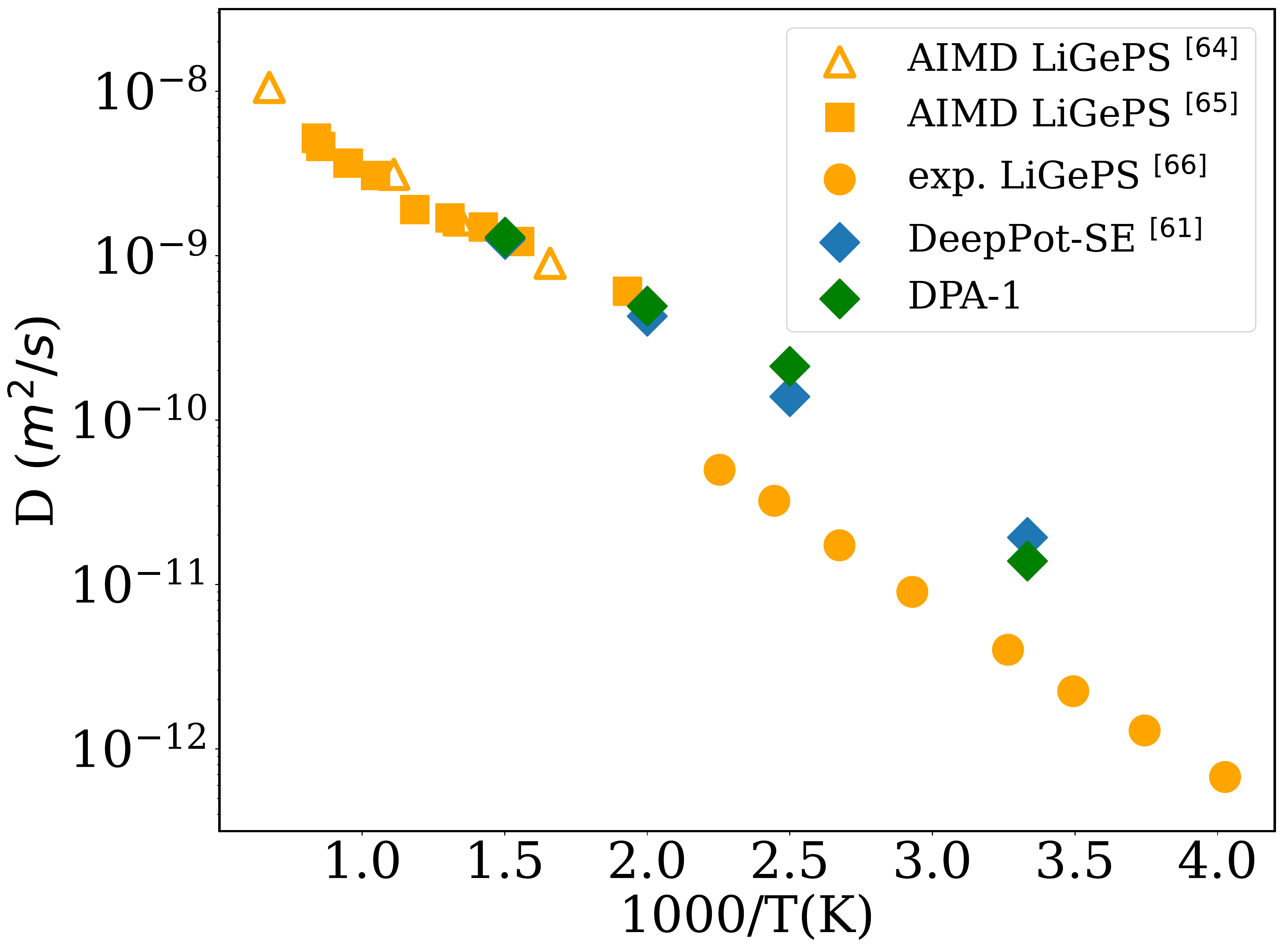}
    \centering
    \caption{  \recheck{
    Diffusion coefficients of Li versus temperature in the \ce{Li10GeP2S12} system obtained from DPA-1 and DeepPot-SE~\cite{huang2021deep} models trained on SSE systems, previous AIMD simulations~\cite{mo2012first, marcolongo2017ionic} and experimental solid-state NMR results~\cite{kuhn2013single}.}}
    \label{fig:diffusion_coefficients}
\end{figure*}
  \recheck{We carried out molecular dynamics simulations on LiGePS systems to assess the diffusion coefficients of Li in relation to temperature, akin to the approach in \cite{huang2021deep}. As illustrated in Fig.~\ref{fig:diffusion_coefficients}, the DPA-1 models trained on SSE systems exhibited a high degree of consistency when compared to AIMD simulations in \cite{mo2012first, marcolongo2017ionic},  experimental studies in \cite{kuhn2013single} and DeepPot-SE simulations in~\cite{huang2021deep}.
% Despite overestimating the diffusion coefficients at room temperature by approximately $5 \times 10^{-12}$m$^2$/s to $15 \times 10^{-12}$m$^2$/s, 
This consistency lends credibility to their reliability.}

\section{  \recheck{Embedded element type interpolation for Nb and Mo in the OC2M dataset}}

\label{Asec:interpolation}
\begin{figure*}
    \includegraphics[width=1.0\textwidth]{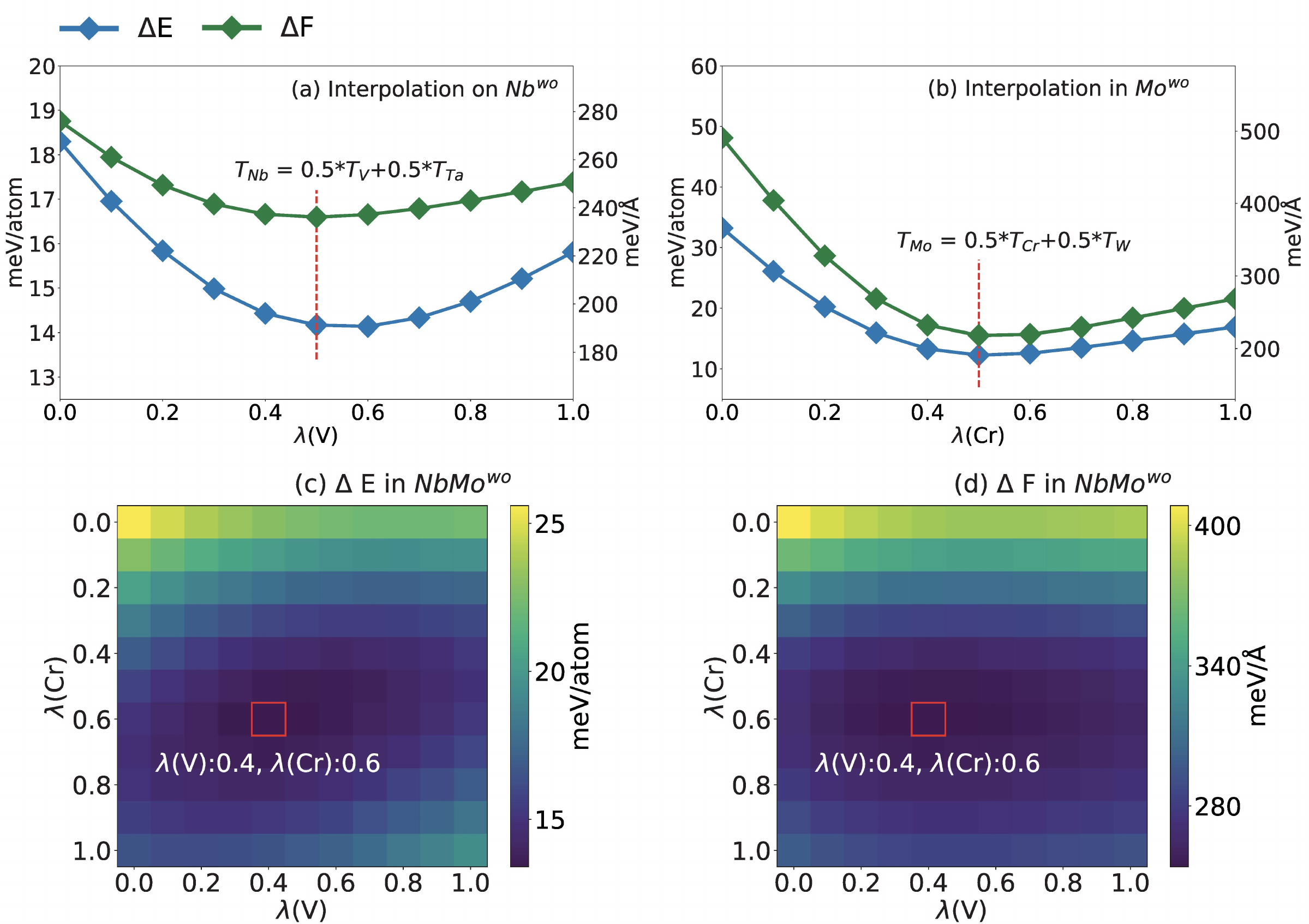}
    \centering
    \caption{
      \recheck{The RMSE of energy and force on the HEA system using embedded element type interpolation for (a)(b) single-element replacements and (c)(d) two-element replacements. {Nb}$^\mathrm{wo}$, {Mo}$^\mathrm{wo}$ and {NbMo}$^\mathrm{wo}$ denote models trained on purposefully altered OC2M datasets that excluded frames containing Nb, Mo, and either Nb or Mo, respectively.}
    }
    \label{fig:interpretability_extra}
\end{figure*}

  \recheck{
To demonstrate the interpretability of DPA-1, we trained models with certain elements omitted from the OC2M dataset and conducted element type interpolation experiments in a manner similar to that described in Sec.~\ref{subsec:interpretability}.
}

\recheck{
In the first two experiments, DPA-1 models named {Nb}$^\mathrm{wo}$, {Mo}$^\mathrm{wo}$ are trained on the OC2M datasets excluding the data frames containing Nb, Mo elements, respectively. 
We used the learned type embeddings of elements from the same periodic table family to interpolate omitted elements as replacements, given by 
\begin{align}\label{eq:nb-inter}
   & T_{Nb} = \lambda\left(V\right)*T_{V} +\left(1-\lambda\left(V\right)\right)*T_{Ta}\\\label{eq:mo-inter}
   & T_{Mo} = \lambda\left(Cr\right)*T_{Cr} +\left(1-\lambda\left(Cr\right)\right)*T_{W},
\end{align}
respectively.
As shown in Fig.~\ref{fig:interpretability_extra}~(a)--(b), we evaluated the RMSE of energy and force on bulk TaNbWMoVAl HEA systems described in Sec.~\ref{dataset_intro}.
% The test system   on HEA systems 
% Three models, named ${Nb}^{wo}$, ${Mo}^{wo}$, and ${NbMo}^{wo}$, were trained on purposefully altered OC2M datasets that excluded Nb, Mo, and both Nb and Mo, respectively.
The {Nb}$^\mathrm{wo}$ model achieves an optimal accuracy of 14.2~meV/atom (energy) and 236~meV/\AA\ (force) at $\lambda\left(V\right) = 0.5$, while the {Mo}$^\mathrm{wo}$ model gives the lowest energy error of 12.3~meV/atom  and force error of 218~meV/\AA\ at $\lambda\left(Cr\right) = 0.5$. 
The optimal accuracy of  {Nb}$^\mathrm{wo}$ and {Mo}$^\mathrm{wo}$ models are almost the same as the baseline errors of 12.0~meV/atom (energy) and 232~meV/\AA\ (force), given by a full OC2M trained DPA-1 model tested on the HEA dataset. 
Surprisingly, the {Mo}$^\mathrm{wo}$ model presents a higher force accuracy than the baseline model.
}

\recheck{
We further remove all data frames containing either Nb or Mo elements, and train a DPA-1 model named {NbMo}$^\mathrm{wo}$. 
The type representation of Nb and Mo are obtained by the interpolation of type embeddings of V, Ta, Cr and W, i.e.~Eqs.~\eqref{eq:nb-inter}--\eqref{eq:mo-inter}.
As shown in Fig.~\ref{fig:interpretability_extra}~(c)--(d), 
the optimal interpolation weights for Nb and Mo are $\lambda\left(V\right) = 0.4$ and $\lambda\left(Cr\right) = 0.6$, respectively.
The corresponding energy error is 13.4~meV/atom lying between the {Nb}$^\mathrm{wo}$ and {Mo}$^\mathrm{wo}$ models.
The force error 253~meV/\AA\ is slightly higher than the  {Nb}$^\mathrm{wo}$ and {Mo}$^\mathrm{wo}$ models, but is only 10\% higher than the baseline. 
The interpolation experiments of Nb and Mo further validate the interpretability of the DPA-1 element type embedding. 
}

%   \recheck{
% We used the learned type embeddings of elements from the same periodic table family to interpolate omitted elements as replacements, given by $T_{Nb} = \lambda\left(V\right)*T_{V} +\left(1-\lambda\left(V\right)\right)*T_{Ta}$ and $T_{Mo} = \lambda\left(Cr\right)*T_{Cr} +\left(1-\lambda\left(Cr\right)\right)*T_{W}$.
% As shown in Fig.~\ref{fig:interpretability_extra}, we evaluated the RMSE of energy and force on bulk TaNbWMoVAl HEA systems in Sec.~\ref{dataset_intro}.
% Notably, both single-element (Fig.~\ref{fig:interpretability_extra}(a)(b)) and two-element interpolation (Fig.~\ref{fig:interpretability_extra}(c)(d)) had minimum RMSEs when interpolation ratios were about 0.5, consistent with the element positions in the periodic table. Since these elements belong to the transition metals group and are closely related, the interpolation curve appears smoother than that of Li.
% }

\section{Results on simple datasets}
\label{Asec:simple datasets}
\begin{table*}[]
    \caption{Prediction RMSE of energies (meV/atom) and atomic forces (meV/\AA) given by DPA-1 (OC2M-pretrained), DPA-1 (from scratch), EANN, and DeepPot-SE. % on validation data of simple bulk systems. 
    DPA-1 (OC2M-pretrained) means that the results were obtained by funetuning a DPA-1 model pretrained using the OC2M dataset.
    %with the same settings as DPM
    %and we also choose smooth edition of DeePMD for comparison.
    For training DPA-1, EANN, and DeepPot-SE, 10\%, 15$\sim$20\%, and 90\% randomly sampled data points are used, respectively. 
    Bold numbers correspond to lowest values.}
    \label{table_simple}
    \centering

\begin{tabular}{cccccccccc}
\hline
\multirow{2}{*}{\textbf{Systems}}                                      & \multirow{2}{*}{\textbf{Sub-systems}}                     & \multicolumn{2}{c}{\textbf{\begin{tabular}[c]{@{}c@{}}DPA-1\\ (OC2M-pretrained)\end{tabular}}}                       & \multicolumn{2}{c}{\textbf{DPA-1}}                                                                                   & \multicolumn{2}{c}{\textbf{EANN}}                                                                                    & \multicolumn{2}{c}{\textbf{DeePot-SE}}                                                                               \\ \cline{3-10} 
                                                                       &                                                           & \begin{tabular}[c]{@{}c@{}}Energy\\ (meV/atom)\end{tabular} & \begin{tabular}[c]{@{}c@{}}Force\\ (meV/\AA)\end{tabular} & \begin{tabular}[c]{@{}c@{}}Energy\\ (meV/atom)\end{tabular} & \begin{tabular}[c]{@{}c@{}}Force\\ (meV/\AA)\end{tabular} & \begin{tabular}[c]{@{}c@{}}Energy\\ (meV/atom)\end{tabular} & \begin{tabular}[c]{@{}c@{}}Force\\ (meV/\AA)\end{tabular} & \begin{tabular}[c]{@{}c@{}}Energy\\ (meV/atom)\end{tabular} & \begin{tabular}[c]{@{}c@{}}Force\\ (meV/\AA)\end{tabular} \\ \hline
Cu                                                                     & \begin{tabular}[c]{@{}c@{}}FCC \\ solid\end{tabular}      & \textbf{0.12}                                              & \textbf{86}                                             & 0.14                                                       & 88                                                      & 0.16                                                       & 89                                                      & 0.18                                                       & 90                                                      \\ \hline
Ge                                                                     & \begin{tabular}[c]{@{}c@{}}diamond \\ solid\end{tabular}  & 0.09                                                       & \textbf{21}                                             & 0.10                                                       & 24                                                      & \textbf{0.07}                                              & 31                                                      & 0.35                                                       & 38                                                      \\ \hline
Si                                                                     & \begin{tabular}[c]{@{}c@{}}diamond \\ solid\end{tabular}  & 0.09                                                       & \textbf{23}                                             & 0.11                                                       & 25                                                      & \textbf{0.08}                                              & 28                                                      & 0.24                                                       & 36                                                      \\ \hline
Al2O3                                                                  & \begin{tabular}[c]{@{}c@{}}Trigonal \\ solid\end{tabular} & \textbf{0.08}                                              & \textbf{40}                                             & 0.08                                                       & 41                                                      & 0.11                                                       & 51                                                      & 0.23                                                       & 49                                                      \\ \hline
\multirow{2}{*}{C5H5N}                                                 & Pyridine-I                                                & \textbf{0.17}                                              & \textbf{21}                                             & 0.21                                                       & 23                                                      & \textbf{0.17}                                              & 41                                                      & 0.38                                                       & 25                                                      \\
                                                                       & Pyridine-II                                               & \textbf{0.24}                                              & \textbf{29}                                             & 0.26                                                       & 32                                                      & 0.35                                                       & 49                                                      & 0.65                                                       & 39                                                      \\ \hline
\multirow{3}{*}{TiO2}                                                  & Rutile                                                    & 0.56                                                       & \textbf{109}                                            & 0.62                                                       & 112                                                     & \textbf{0.51}                                              & 133                                                     & 0.96                                                       & 137                                                     \\
                                                                       & Anatase                                                   & \textbf{0.87}                                              & \textbf{144}                                            & 0.91                                                       & 151                                                     & 1.03                                                       & 183                                                     & 1.78                                                       & 181                                                     \\
                                                                       & Brookite                                                  & \textbf{0.50}                                              & \textbf{88}                                             & 0.52                                                       & 91                                                      & 0.55                                                       & 96                                                      & 0.59                                                       & 94                                                      \\ \hline
\multirow{5}{*}{\begin{tabular}[c]{@{}c@{}}MoS2\\ +Pt\end{tabular}}    & MoS2 slab                                                 & \textbf{0.14}                                              & \textbf{11}                                             & 0.17                                                       & 12                                                      & 0.26                                                       & 19                                                      & 5.26                                                       & 23                                                      \\
                                                                       & bulk Pt                                                   & \textbf{0.25}                                              & \textbf{41}                                             & 0.26                                                       & 43                                                      & 0.38                                                       & 64                                                      & 2.00                                                       & 84                                                      \\
                                                                       & Pt surface                                                & \textbf{0.59}                                              & \textbf{45}                                             & 0.68                                                       & \textbf{45}                                             & 3.8                                                        & 86                                                      & 6.77                                                       & 105                                                     \\
                                                                       & Pt cluster                                                & \textbf{1.21}                                              & \textbf{41}                                             & 1.33                                                       & \textbf{41}                                             & 9.7                                                        & 152                                                     & 30.6                                                       & 201                                                     \\
                                                                       & Pt on MoS2                                                & 4.43                                                       & \textbf{86}                                             & 5.02                                                       & 88                                                      & \textbf{1.33}                                              & 91                                                      & 2.62                                                       & 94                                                      \\ \hline
\multirow{2}{*}{\begin{tabular}[c]{@{}c@{}}CoCrFe\\ MnNi\end{tabular}} & rand.occ.I                                                & \textbf{1.31}                                              & \textbf{367}                                            & 1.43                                                       & 387                                                     & 2.3                                                        & 410                                                     & 1.68                                                       & 394                                                     \\
                                                                       & rand.occ.II                                               & \textbf{1.74}                                              & \textbf{381}                                            & 1.93                                                       & 389                                                     & 3.3                                                        & 415                                                     & 5.29                                                       & 410                                                     \\ \hline
\end{tabular}
\end{table*}
As shown in Table \ref{table_simple}, we trained DPA-1 from scratch on \textbf{simple bulk systems~\cite{zhang2018deep}} and compared with the embedded atom neural network potential (EANN)~\cite{zhang2019embedded} and DeepPot-SE. 

This small dataset contains two types of systems. The first type includes general systems, such as relatively easy \ce{Cu}, \ce{Ge}, \ce{Si}, \ce{Al_2O_3} with one single solid phase, and more challenging systems like \ce{C_5H_5N} (pyridine) and \ce{TiO_2} with two and three phases, respectively. 
The second type of systems contains a grand-canonical-like system of supported \ce{Pt} clusters on a \ce{MoS_2} slab and a \ce{CoCrFeMnNi} high entropy alloy (HEA) system. 

For training DPA-1 and EANN, only 10\% and 15$\sim$20\% randomly selected data points are used respectively, while DeepPot-SE used 90\%.
The results show that, even with less training samples and trained from scratch, DPA-1 still mostly outperforms the other methodologies, especially in terms of force prediction accuracy. Interestingly, all the finetuned model based on OC2M-pretrained outperforms other models.

\section{Hyperparameters in sample efficiency tests}
To generate the learning curves of sample efficiency in Sec.~\ref{subsec:sample efficiency}, we started from 50 randomly selected samples and set the training steps to be 100k during each iteration. The probabilities to visit the historical and newly added samples are set to 0.9 and 0.1, respectively.

\end{document}